\documentclass[a4paper,conference]{IEEEtran}
\IEEEoverridecommandlockouts
\pdfoutput=1    

\usepackage{xcolor}
\definecolor{wong-black}        {HTML}{000000}
\definecolor{wong-lightorange}  {HTML}{E69F00}
\definecolor{wong-lightblue}    {HTML}{56B4E9}
\definecolor{wong-green}        {HTML}{009E73}
\definecolor{wong-yellow}       {HTML}{F0E442}
\definecolor{wong-darkblue}     {HTML}{0072B2}
\definecolor{wong-darkorange}   {HTML}{D55E00}
\definecolor{wong-pink}         {HTML}{CC79A7}

\usepackage[accsupp]{axessibility}  

\usepackage[hyphens]{url}
\usepackage[bookmarks=false]{hyperref}

\usepackage{cite}
\usepackage{amsmath,amssymb,amsfonts}
\usepackage{algorithmic}
\usepackage{graphicx}
\usepackage{booktabs}
\usepackage{textcomp}
\usepackage[nolist, nohyperlinks, printonlyused]{acronym} 

\def\BibTeX{{\rm B\kern-.05em{\sc i\kern-.025em b}\kern-.08em
    T\kern-.1667em\lower.7ex\hbox{E}\kern-.125emX}}

\usepackage{orcidlink}

\usepackage[per-mode=symbol]{siunitx}
\DeclareSIUnit \beli {Bi}  
\DeclareSIUnit \belm {Bm}  

\usepackage{svg}
\usepackage{comment}
\usepackage[utf8]{inputenc}
\DeclareUnicodeCharacter{1D700}{$\epsilon$}
\DeclareUnicodeCharacter{03B5}{$\epsilon$}
\DeclareUnicodeCharacter{03B8}{$\theta$}

\usepackage[caption=false]{subfig}

\usepackage[english]{babel}
\usepackage [autostyle, english = american]{csquotes}

\begin{document}


\title{6G Satellite Direct-to-Cell Connectivity: “To distribute, or not to distribute, that is the question"}


\author{\IEEEauthorblockN{Diego Tuzi\,\orcidlink{0000-0001-7477-9473}\IEEEauthorrefmark{1},
Thomas Delamotte\,\orcidlink{0000-0003-4313-3033}\IEEEauthorrefmark{1} and
Andreas Knopp\,\orcidlink{0000-0001-7798-0535}\IEEEauthorrefmark{1}}
\IEEEauthorblockA{\IEEEauthorrefmark{1}\textit{Institute of Information Technology, University of the Bundeswehr Munich,  85579 Neubiberg Germany}\\
Email: paper.sp@unibw.de, \{name.surname\}@unibw.de}
}

\IEEEaftertitletext{\vspace{0em}\noindent\begin{minipage}{\textwidth}
\centering
\small
© 2023 The authors. This paper was presented at the ESA SatNEx School 2023 workshop \textit{“Satellite 6G: Challenges and Solutions”}, University of Siena, Italy (April 18–20, 2023). It received the Best Idea Award and has not been peer-reviewed.
\end{minipage}\vspace{0.5em}}

\maketitle




\begin{acronym}
    \acro{ntn}[NTN]{non-terrestrial network}
	\acro{leo}[LEO]{low Earth orbit}
	\acro{vsat}[VSAT]{very small aperture terminal}
	\acro{d2c}[D2C]{direct-to-cell}
    \acro{tn}[TN]{terrestrial network}
    \acro{ue}[UE]{user equipment}
    \acro{hh}[HH]{handheld}
    \acro{geo}[GEO]{geostationary Earth orbit}
    \acro{meo}[MEO]{medium Earth orbit}
    \acro{ff}[FF]{formation flying}
    \acro{dss}[DSS]{distributed satellite system}
    \acro{gl}[GL]{grating lobe}
    \acro{uav}[UAV]{unmanned aerial vehicles}
    \acro{hpbw}[HPBW]{half power beam width}
    \acro{elsa}[ELSA]{enhanced logarithmic spiral array}
\end{acronym}


\begin{abstract}
Direct-to-cell connectivity between satellites and common terrestrial handheld devices represents an essential feature of 6G. The industry is considering different type of constellations but using classical single satellite solutions based on phased array antennas. This article proposes to decompose a classical single satellite into a swarm of multiple small platforms (e.g. CubeSats) each equipped with one or a small number of radiating elements. The platforms are spaced far apart to create a large virtual aperture. The use of small satellites promises cost reduction for production and launch, while the distributed nature of the system introduces interesting features, such as scalability and fault tolerance. This perspective article provides insights into the opportunities and a discussion of the research challenges for the feasibility of the proposed approach.
\end{abstract}




\section{Introduction}
\label{sec:introduction}
One of the most interesting uses of satellite networks in 6G is to provide direct connectivity to terrestrial \ac{ue}. In the literature, the term direct connectivity generally addresses two main scenarios depending on the considered macro-category of \ac{ue}. The first type of direct connectivity considers \acp{vsat}. A \ac{vsat} can be a parabolic dish of around \SI{60}{\centi\metre} in diameter, similar to \ac{geo} terminals, but with a motorized system to track the \ac{leo} satellites. A \ac{vsat} can also be a phased array antenna able to electronically steer the main beam and track the satellite. \acp{vsat} have good performance in terms of power and antenna capabilities, hence, they can use high frequency (mainly Ku and Ka bands). Famous examples of operational/planned constellations for \acp{vsat} are Starlink (SpaceX), OneWeb, and Project Kuiper (Amazon). 

\begin{figure*}
\centering
\includegraphics[width=\textwidth]{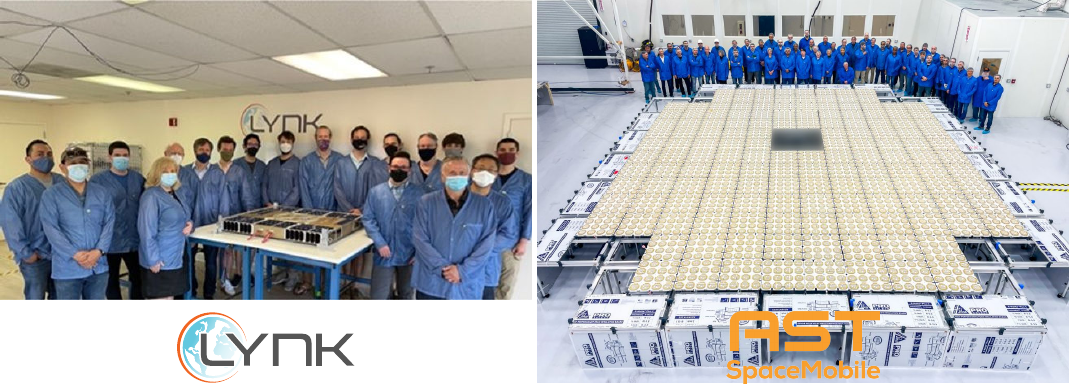}
\caption{Recent satellites developed for \ac{d2c} connectivity: Lynk Global \cite{lynkglobal} is considering smaller satellites compared to AST SpaceMobile \cite{astspacemobile}.}
\label{fig:D2C_companies}
\end{figure*}

The second type of direct connectivity mainly considers common terrestrial \ac{hh} devices referred to as \ac{d2c} connectivity. \acp{hh} are for example common smartphones characterized by low power and low antenna performance that can usually connect to terrestrial base stations in their vicinity. A high-speed connectivity with state-of-the-art satellite systems is not possible even though lower frequencies (UHF, L- and S-bands) and low Earth orbits enable to reduce the impact of atmospheric attenuation and path loss. Therefore, the space segment needs to be entirely rethought to offer higher transmit gains and optimize the reuse of the spectral resources. Even in this case, there are famous examples from the industry, such as satellites developed by Lynk Global and AST SpaceMobile (Fig. \ref{fig:D2C_companies}). The former is planning a mega constellation of around 5110 satellites with a satellite size of around \SI{4}{\metre\squared} while the latter is planning a constellation of around 170 satellites with a massive satellite with a size of around \SI{128}{\metre\squared} \cite{laursen_your_2022}. 
This paper focuses on the \ac{d2c} use case and considers one single satellite of a \ac{leo} constellation. In particular, considering the two previous approaches from the industry there are considering different approaches to realize the constellation. Despite the differences, they are both considering classical solutions based on phased array antennas where the radiating elements are usually organized in a regular lattice (e.g. rectangular) spaced with a uniform distance, usually around half the wavelength. 

This paper proposes an alternative solution based on a \ac{dss}. The main idea is to decompose a single satellite equipped with a phased array antenna in a satellite swarm configuration of multiple small platforms (e.g. CubeSats) each equipped with one or a small number of radiating elements. The platforms are spaced much more than half the wavelength to create a large virtual aperture achieving high antenna performance. The platforms of the swarm are connected using a wireless or wired connection and coordinated to achieve the coherent transmission and reception of signals. 
The use of small satellites promises cost reduction for production and launch of the swarm compared to a classical single satellite. In addition, the distributed nature of the system provides interesting features, such as scalability and fault tolerance. 

This perspective article elaborates on the results in \cite{10068542} by discussing the main opportunities and challenges for use of swarms in the \ac{d2c} use case. The ultimate goal is to provide elements to address the Hamletic doubt in the title.

\section{Satellite swarms for \ac{d2c} connectivity}
\label{sec:method}

Swarms for D2C are based on two main concepts, phased array antennas, and \acp{dss}. 

A phased array antenna consists of a group of radiating elements connected via internal circuits and organized in a defined geometry, capable of changing the shape and direction of the radiation pattern without physically moving the antenna. The array elements are usually organized in a regular lattice (e.g. rectangular, circular) spaced with a uniform distance, usually around half the wavelength. The radiating elements transmit a phase-shifted version of the same signal, where phase shifts are calculated to provide a constructive summation of the signals in the desired direction. The resulting radiation pattern offers better gain, directivity, and performance in a given direction than the single-element radiation pattern. Classical single satellite solutions for \ac{d2c} connectivity are considering solutions based on phased array antennas. Fig. \ref{fig:D2C_companies} shows satellites developed by Lynk Global and AST SpaceMobile, certainly two different approaches in terms of size.

On the other hand, a \ac{dss} is composed of multiple satellites that are coordinated with each other. In the context of \ac{dss}, the satellite swarm refers to a configuration where a multitude of identical and autonomous satellites can achieve a common goal with their common behavior. The swarm configuration is also closely related to other terms, such as fractionated systems and \ac{ff}. In fractionated systems, the different satellites of the same \ac{dss} have different functions, while the term \ac{ff} refers to the problem of maintaining a desired separation, orientation, or relative position between satellites belonging to the same multi-satellite configuration. In addition, the architecture is a distributed phased array antenna from the antenna point of view. The architecture discussed in this article is a hybrid satellite swarm and fractionated \ac{dss} configuration, organized in a \ac{ff} creating a distributed phased array antenna, that for the sake of simplicity is referred to as swarm.

\begin{figure*}[!t]
\centering
\includegraphics[width=\textwidth]{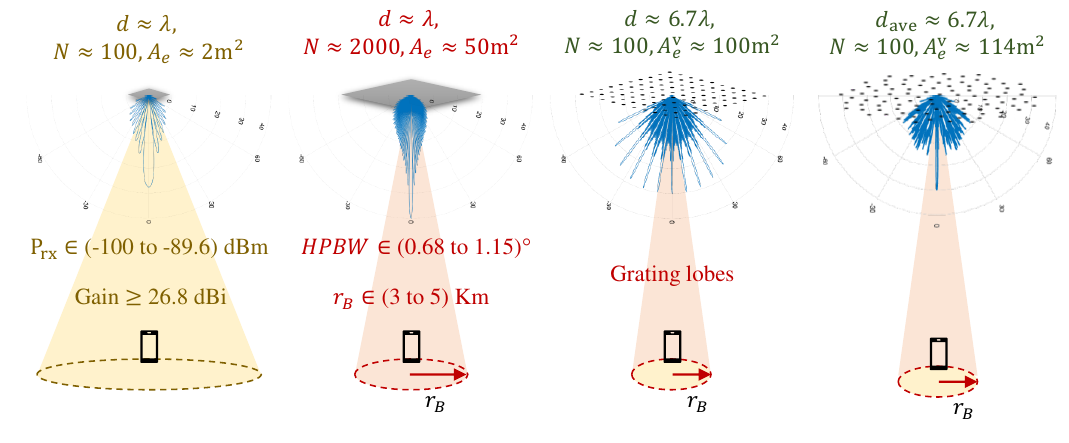}
\caption{From a classical single satellite with phased array antennas to a swarm of multiple CubeSats. Summary of the swarm design in \cite{10068542}.}
\label{fig:results}
\end{figure*}

Swarms for \ac{d2c} decompose the classical single satellite with a phased antenna with $N$ elements in $N_p$ platforms with $N_r$ radiating elements. The number of radiating elements ($N_r$) can be equal to one or more, according to the space limit of CubeSats. Clearly, if $N_r=1 \implies N_p=N$. Swarms usually have one or more satellites with enhanced capabilities, called the leader(s) or chief(s), while the other satellites are called followers or deputies. The satellites of the swarm have an average inter-element distance ($d_{ave}$) much higher than half the wavelength to create a large virtual antenna aperture ($A_e^v$).

Although satellite swarms could appear far from reality, numerous research and space flight demonstrations have been conducted on swarms for different fields of application like astronomy, deep-space communications, meteorology, and environmental uses \cite{8293791}. Most missions have been conducted with a limited number of satellites, but in recent decades research has been considering systems with an increasing number of satellites, especially with the advent of small satellites \cite{Bandyopadhyay2015,9977426}. A recent example of commercial flying technology is the Cluster 1 and 2 solutions from HawkEye 360 for radio signal mapping application. HawkEye 360's satellites fly in dense formations in groups of three to determine the location of the RF signal source \cite{HawkEye}. Swarms have also been considered in \ac{uav} \cite{6162964}. The use of a swarm of thousands of independent tiles is also considered for beaming solar power to Earth from space \cite{SpaceSolar}.

Nevertheless, swarms of small satellites for telecommunication purposes have received limited attention. The recent paper in \cite{Bacci_2023} presented a formation of sub-arrays in \ac{geo} and \ac{leo} scenarios, where a large array was subdivided in smaller sub-arrays in a squared formation, resulting in \ac{leo} performance similar to the classical single antenna solution. In contrast, the paper in \cite{10068542}, from the same authors of this article, presented a swarm-based antenna array for \ac{d2c} connectivity, showing that through the control of the geometry, swarms can bring several opportunities in the \ac{d2c} use case. The advantages of a distributed approach for \ac{d2c} connectivity are also recognized in the AST SpaceMobile patent from \cite{US9973266B1}, but without addressing the challenges.

\subsection{Antenna performance}
Swarms can achieve a large virtual aperture by increasing the distance between the platforms. As known from the antenna array theory, increasing the distance between the radiating elements more than a specific threshold leads to the \ac{gl} problem. By carefully designing the geometry of the swarm, the \acp{gl} can be mitigated. The paper in \cite{10068542} proposed a family of geometry named \ac{elsa} based on \cite{boeringer_phased_2002,vigano_sunflower_2009} that can mitigate the \ac{gl} while maintaining a simple mathematical formulation for the position of the platforms in the formation. A summary of the design process carried out in \cite{10068542} is in Fig. \ref{fig:results}. 
Starting from the left of Fig. \ref{fig:results}, a mid-size satellite at \ac{leo} altitude of \SI{500}{\kilo\metre} with hundreds of radiating elements can fulfill the power requirements ($P_{rx}$) of a common handheld terminal. On the other hand, a satellite requires thousands of elements to reduce the \ac{hpbw} and achieve an Earth coverage radius ($r_B$) of less than \SI{5}{\kilo\metre}. A squared swarm of small satellites with an increased distance ($d$) between the platforms can provide the required power level and the required cell size, but generates several minor lobes with an energy level comparable to the main lobe one (\acp{gl}). A swarm with a \ac{elsa} geometry can maintain the performance of the main lobe, but it mitigates the \acp{gl}. The resulting beam provides the required energy level and coverage, reducing the number of radiating elements by an order of magnitude.

\subsection{Cost reduction opportunities}
Swarms offer a twofold cost reduction opportunity. Classical single satellites have a unique design and production chain, whereas swarms are based on CubeSats and commercial off-the-shelf products that can be easily assembled. This can lead to reduced production costs. Furthermore, a CubeSat platform can integrate all components, including solar panels, in a lightweight cubic shape. Therefore, CubeSats can reduce the total weight and can be flexibly arranged in the rocket for launch, resulting in reduced launch costs. 

\subsection{Distributed features}
Swarms inherit the advantages of \acp{dss}. In particular, the workload distribution on multiple elements makes the system fault-tolerant: a failure of single or multiple elements leads to a graceful performance degradation but not to an interruption of service. \acp{dss} also introduce scalability: the performance of swarms can be controlled by the number of platforms and the distance between them. A swarm could be launched in an initial configuration to have a larger beam considering a small size constellation. Adding other swarms to the constellation, previous swarms could be reconfigured with a larger inter-platform distance to reduce the beam. In addition, a swarm could be deployed with a reduced number of platforms, and additional launches could increase the number of platforms to increase performance.

\section{Research challenges}
\label{sec:evaluation}

Despite the benefits, swarms present several difficult challenges to deal with.

\subsection{Multi-beam coverage optimization}
Although the control of the geometry mitigates the \acp{gl}, the resulting beam pattern maintains an average level of interference outside the main lobe of around \SI{-20}{\decibel}. Classical phased array antenna solutions can reduce the level of interference outside the main lobe by simply applying tapering techniques via known windows. Preliminary results on the application of known tapering techniques to \ac{elsa} geometries do not produce the same results. For this reason, the generation of multiple beams has to deal with a higher level of inter-beam interference. In addition, narrowing the beams could reduce the coverage of the swarm impacting the size of the whole constellation.

\subsection{Formation Flying stability}
Swarms, organized according to an optimized geometry, must be kept stable during the flight around the Earth. However, various effects like Earth’s oblateness, atmospheric drag, and solar radiation pressure significantly influence the positions of the elements, making periodic orbit corrections necessary to maintain stability. Promising studies on \ac{ff} \cite{morgan_swarm-keeping_2012,9438513}, electric propulsion, electromagnetic forces, and the growing interest of the scientific community and space agencies \cite{G002868, sanchez_starling1_2018, esa_join_nodate} could lead to promising developments in the near future. The impact of \ac{ff} stability on the beamforming results has to be quantified and opportunely mitigated.

\subsection{Synchronization}
Although \ac{ff} stability is important, a large degradation of the performance when the optimized geometry varies from the perfect one is not expected, since the grating lobes are mitigated by breaking the periodicity of regular geometries. Most importantly, the phase differences between the satellites and a common reference point must be estimated with a certain degree of accuracy, and the phase shifts updated accordingly. For this reason, synchronization is one of the most important aspects of swarms. The scientific literature offers multiple solutions to approach the problem \cite{9492307}. According to a preliminary analysis, a synchronization process based on open-loop strategies, RF, and differential GPS technologies could be sufficient to achieve the required level of coordination, but the precise impact of the imperfect synchronization on the performance needs to be investigated.

\subsection{System design aspects}
In addition to the research challenges described above, several design aspects influence the level of complexity of the entire system, such as the division of tasks between the leader(s) and followers of the same swarm, or the division of tasks between space and ground sections.
Furthermore, it must be emphasized that synchronization and \ac{ff} stability are stringent requirements mainly for free-flying systems. Systems with a wired connection between satellites (tethered), if feasible, could drastically reduce these requirements. Tethered systems could also use very small satellite platforms based on the "satellite on a chip" or "satellite on a printed circuit board" concept to further reduce the production and launch costs.

\section{Conclusion}
\label{sec:conclusion}

This article presented swarms for satellite \acl{d2c} connectivity, an alternative solution to the classical single satellite paradigm. This new approach promises increased antenna performance, cost reductions, and interesting distributed features, but a research effort into the identified challenges is needed to clarify the Hamletic doubt in the title.

\section{Acknowledgment}
\label{sec:acknowledgment}
Part of this research paper was supported by European Space Agency (ESA) within the project SatNExV WI Y2.2-A under the Grant No: 4000130962/20/NL/NL/FE (the opinions and conclusions presented herein are those of the authors and can in no way be taken to reflect the official opinion of the European Space Agency). Part of this research paper was funded in part by dtec.bw – Digitalization and Technology Research Center of the Bundeswehr. dtec.bw is funded by the European Union – NextGenerationEU.


\bibliographystyle{IEEEtran}
\bibliography{references}

\end{document}